\documentclass[12pt]{article}

\parskip=\medskipamount            % space between paragraphs (LaTeX)
\def\7#1#2{\mathop{\null#2}\limits^{#1}}        % puts #1 atop #2

\def\beee{\begin{equation}}
\def\eeee{\end{equation}}

\begin{document}
\bibliographystyle{unsrt}
 
\begin{center}
\textbf{The Parton Model}\\
[5mm]
O.W. Greenberg\footnote{email address, owgreen@physics.edu.}\\
{\it Center for Fundamental Physics\\
Department of Physics\\
University of Maryland\\
College Park, MD~~20742-4111, USA,\\
}
%University of Maryland Preprint PP-}\\
~\\
\end{center}

\begin{abstract}
I give a brief review of the parton model.
\end{abstract}                                          

The parton model pictures hadrons as a collection of pointlike quasi-free particles. 
The model describes the cross section for
high-energy scattering of hadrons
with another particle as an incoherent sum of the cross sections of the 
pointlike partons
in the hadron with the other particle. The hadronic factors in the
cross sections are parametrized by
``structure functions.'' The parton model expresses the structure functions
in terms of parton distribution functions that give the longitudinal momentum
distribution of the partons in the given
hadron. The parton distribution functions are found from experimental 
data in a given process
and are used in the description of other processes. 

The prototype process for the
parton model is $e N \rightarrow e^{\prime} X$, where $e$ and $e^{\prime}$ are
the incident and scattered electron, $N$ is the target nucleon, and $X$ is the
set of final state hadrons. The particles in the 
final state $X$ are not measured, so the
cross section is for the sum over all hadronic final states, an ``inclusive''
cross section. This contrasts with an ``exclusive'' cross section in which the final
states are restricted to a specific subset. In the prototype process, 
$eN \rightarrow e^{\prime}X$, the kinematics of the inclusive scattering
depends on the momentum transfer $q=k-k^{\prime}$ from the electron to the hadrons
and the invariant mass, $W$, of the hadronic final state, where 
$W^2=(p+q)^2=M^2+2M\nu+q^2$, and $M$ is the mass of the target nucleon or other
hadron.
Here $k$ and $k^{\prime}$ are the energy-momentum 4-vectors of the incident and scattered
electron, $p$ is the energy-momentum 4-vector of the target hadron, and $\nu=E-E^{\prime}$
is the energy transfer to the target hadron in its rest frame. 

J.D. Bjorken~\cite{bj0} predicted that
the hadronic factor in the cross section would depend only on the ratio
$x=(-q^2)/(2 p \cdot q)=(-q^2)/(2M\nu)$, rather than on $\nu$ and $-q^2$
separately, on the basis of an algebra of local currents. 
This property, called ``scaling,'' was expected to hold in the
``deep inelastic'' limit in which the energy transfer and momentum transfer
are much larger than the target hadron mass. 
R.P. Feynman~\cite{fey} interpreted scaling in terms of constituents of the nucleon that
he called ``partons.'' Bjorken and E.A. Paschos~\cite{bj1,bj2}
gave early discussions of electron-nucleon and neutrino-nucleon scattering
in the deep inelastic limit.
The Bjorken $x$ can be identified
with the fraction of the longitudinal hadron momentum carried by a given parton. 

C.G. Callan and D.J.
Gross~\cite{cal} showed that the commutators of the electric current give information about
the carriers of electric charge. Subsequent data on deep inelastic scattering showed that
the carriers of charge have spin 1/2 and can be identified with quarks (see~\cite{nob} 
for early data and~\cite{cteq, blu} for recent data in
the references).
Other sum rules together with data show that the charged partons carry only about
1/2 of the energy-momentum of the nucleon. The other half is carried by gluons and other neutral
particles. Several reviews discuss sum rules below (see A.J. Buras~\cite{bur},
C. Bourrely and J. Soffer~\cite{bou} and F. Close~\cite{clo} in the references). 

Surprisingly, scaling sets in at
rather low energy and momentum transfer, so-called ``precocious'' scaling.~\cite{blo}
The paper of Bloom and Gilman also pointed out a duality between resonances and
smooth scaling behavior which later led to the dual resonance model and even later
to string theory. 
The partons are identified with the ``valence'' quarks that account for the
electric charge, isospin and strangeness of the hadron, and with ``sea'' quarks
that correspond to extra quark-antiquark pairs as well as with ``gluons,'' which
are quanta of the color gauge group that mediate quark interactions and
have zero electric charge, isospin and strangeness.     
S.D. Drell, D.J. Levy and T.-M. Yan extended the parton model to hadron-hadron
scattering and gave the celebrated Drell-Yan mechanism for the production of lepton
pairs (see~\cite{dre} in the references for a review).

More detailed processes, such as semi-inclusive
processes in which some of the final state hadrons are measured, require parton
fragmentation functions~\cite{fey2}, as well as parton distribution functions, for their
description. The fragmentation functions account for the conversion of partons
into hadrons in the final states. Gross and Wilczek~\cite{gro2} and H. Georgi and
P{olitzer~\cite{geo} showed that quantum chromodynamics predicts logarithmic corrections
to scaling. The DGLAP formalism~\cite{dglap} expresses these corrections in parton
language.

Scattering experiments with polarized beams and targets give information
that cannot be obtained from unpolarized experiments. The EMC experiment with
polarized muons scattering on polarized protons~\cite{emc} led to the ``spin crisis,''
that only about 1/4 of the spin of the proton is carried by quarks~\cite{kar} 
(see reviews in~\cite{ji}).

Feynman gave arguments that partons don't interact with each other in first 
approximation because in the limiting infinite momentum frame there is a separation
of scales between the (slow) parton-parton interactions and the (fast) 
interaction with the scattered lepton.~\cite{fey2} The 
running of coupling constants that follows from asymptotic freedom provides 
further understanding of
the mystery that quarks are permanently confined in hadrons viewed at low energy,
but are quasi-free viewed as partons at high energy.~\cite{growil, pol}

R.E. Taylor, H.W. Kendall and J.I. Friedman describe the pathbreaking experimental 
discoveries that stimulated the invention of the
parton model~\cite{nob}. P.M. Nadolsky, et. al.~\cite{cteq}  and J. Blumlein, et. al.~\cite{blu}
analyse recent data on parton distributions. 

Note: References [1] through [21] are primary references. References [22] through [25]
are secondary references.

\end{document}